\documentclass[12pt,journal,compsoc]{IEEEtran}
\ifCLASSOPTIONcompsoc
\else
\fi
\ifCLASSINFOpdf
\else
\fi
\usepackage{graphicx}
\usepackage[justification=centering]{caption}

\renewcommand{\baselinestretch}{0.93}

\begin{document}
%
\title{Modeling context and situations in pervasive computing environments}
%
%
%
%

\author{Preeti~Bhargava, Shivsubramani~Krishnamoorthy, Ashok~Agrawala\\
Department of Computer Science, University of Maryland, College Park
E-mail:{prbharga,shiv,agrawala}@cs.umd.edu
}

\IEEEcompsoctitleabstractindextext{%
\begin{abstract}
In pervasive computing environments, various entities often have to cooperate and integrate seamlessly in a \emph{situation} which can, thus, be considered as an amalgamation of the context of several entities interacting and coordinating with each other, and often performing one or more activities. However, none of the existing context models and ontologies address situation modeling. In this paper, we describe the design, structure and implementation of a generic, flexible and extensible context ontology called Rover Context Model Ontology (RoCoMO) for context and situation modeling in pervasive computing systems and environments. We highlight several limitations of the existing context models and ontologies, such as lack of provision for provenance, traceability, quality of context, multiple representation of contextual information, as well as support for security, privacy and interoperability, and explain how we are addressing these limitations in our approach. We also illustrate the applicability and utility of RoCoMO using a practical and extensive case study.
\end{abstract}

\begin{IEEEkeywords}
Context-aware Computing, Context Modeling and Representation, Situation Modeling
\end{IEEEkeywords}}

\maketitle

\IEEEdisplaynotcompsoctitleabstractindextext

%
\IEEEpeerreviewmaketitle

\section{Introduction}

Recent years have witnessed rapid advances in enabling technologies for pervasive computing environments - an important step being context-awareness in systems. Dey and Abowd \cite{abowd1999towards} describe a context-aware system as one that ``uses context to provide relevant information and/or services to the user, where relevancy depends on the user's task." Context awareness enables a new class of applications in pervasive computing that can help users navigate through unfamiliar territory, find preferred restaurants nearby, receive messages in the least obtrusive manner, get extra sleep when meetings are canceled, find people with similar interests, and so on. The use of context information in these applications reduces the amount of human effort and attention that an application needs to service the user's requests.

Moreover, in pervasive computing environments, various entities often have to cooperate and integrate seamlessly in a \emph{Situation} to achieve a common objective. Thus, Situation Awareness can be defined as ``the capability of the entities in pervasive computing environments to be aware of situation changes and automatically adapt themselves to such changes to satisfy user requirements, including security and privacy."  and a \emph{Situation} can be described as ``a set of contexts in the application over a period of time that affects future system behavior." \cite{yau2006hierarchical}. A situation can be considered as an amalgamation of the context of several entities interacting and coordinating with each other, and often performing one or more activities.

The context model forms the underlying framework for modeling and representing context in the pervasive computing environment and context-aware systems. To support context- and situation-awareness, and adaptation of the entities in pervasive computing environments, it is necessary to model and specify context and situations in a suitable way such that the contextual information can be easily exchanged, shared and reused. As discussed in several papers including Chen et al.  \cite{chen2004soupa} and Krishnamoorthy et al. \cite{krishnamoorthycj}, ontologies are a powerful tool for modeling context and the encompassing situations in context-aware systems because they promote knowledge sharing and reuse across different applications and services interacting in a pervasive computing environment, thus, enhancing their interoperability. They allow context-aware systems to use existing logic reasoning mechanisms to deduce high-level, conceptual context from low-level, raw context, and handle uncertainty and inconsistency in context. They can be combined to form a more complex ontology and save the effort. 

However, none of the existing context models and ontologies address situation modeling in a dynamic environment where the situation constantly evolves. To address this limitation, we described the design of a general and intelligent context-aware middleware called Rover II and its general, flexible and extensible context model for context and situation modeling, called Rover Context Model (RoCoM), in Krishnamoorthy et al. \cite{krishnamoorthycj} and Bhargava et al. \cite{bhargava2012ontological}. RoCoM has four Primitives - Entity, Event, Activity and Relationship. These Primitives are the building blocks of every context-aware system or middleware built using this model. Any piece of contextual information in the system can be attached to one of these primitives and any situation can be modeled via them. 

We introduced the Rover Context Model Ontology (RoCoMO), which is the underlying ontology for RoCoM and is currently deployed and implemented in Rover II, in \cite{bhargava2012ontological}. In this paper, we describe its design, structure and implementation in detail. Each primitive of RoCoM corresponds to a top level concept in RoCoMO from which other concepts are derived. 
Our main contributions in this paper are:
\begin{itemize}
\item Highlighting several shortcomings of other existing standard models and ontologies for context and situation modeling and demonstrating the utility of RoCoMO's capabilities that address those shortcomings,
\item Illustrating the benefits, applicability and utility of RoCoMO, as opposed to other existing models and ontologies, using a simple and practical case study for context and situation modeling in pervasive computing environments.
\end{itemize}

The rest of this paper is organized as follows. In Section \ref{sec:scenario}, we describe a case study which we use in this paper to motivate and illustrate the benefits and utility of RoCoMO. We briefly discuss the existing approaches to context and situation modeling, and highlight their limitations in Section \ref{sec:related}. In Section \ref{sec:design}, we explain the design and implementation of RoCoMO and how it addresses several limitations of contemporary approaches such as lack of provision for provenance, quality of context, multiple representations of contextual information as well as support for security etc. We illustrate the benefits of RoCoMO by revisiting the case study in Section \ref{sec:usecases}. We conclude and outline future work in Section \ref{sec:conclusion}.

\section{Motivation\label{sec:scenario}}

We describe a simple but practical case study here in order to motivate and illustrate the varied nature of context, and the capabilities that context models and ontologies should possess for representing and modeling this situation in real time pervasive computing systems and environments. We will return to this case study in Sections \ref{sec:evaluation} and \ref{sec:usecases} to illustrate RoCoMO's modeling capabilities. For illustration, we have selected a situation from the domain of rescue and evacuation but this, by no means, restricts RoCoMO's applicability and generality.

\emph{A fire incident takes place in a room on the fourth floor of a building on a university campus. Fire fighters and responders are using a context-aware system to coordinate the rescue efforts. A responder, using the system, gets updated readings from two temperature sensors in the room on fire. Using this contextual information, the system determines the time that responders have to evacuate the building before the whole building is engulfed in flames. The temperature information also has a quality measure attached to it to convey any inaccurate or incomplete information.
Another responder is accessing the system to get confidential floor maps of the building and also determine the evacuation route people should take based on the floormaps and the time remaining.}

Representing and modeling this situation in a context-aware system (such as Rover II), using the existing ontologies, is not trivial. It involves interaction between several entities such as responders which perform one or more activities. The entire situation is catalyzed by an event like the fire incident and the goal of the situation is to evacuate the building. Each activity, whether being performed by the system or an entity, is driven by the goal or a sub-goal. Some of the activities can occur simultaneously, for instance, calculating the time remaining for evacuation and accessing the floor maps of the building. Other activities need to be performed sequentially - the evacuation routes can be determined only when the floor maps are available. The room has contextual information, that needs to be modeled and represented, such as the temperature readings of the room. To avoid ambiguity, the information must be clearly marked with its source (which sensor it is coming from) as well as the encoding format (whether it is in Celsius or Fahrenheit). This requires support for encoding format as well as provenance. The model should also support attachment of quality attributes to contextual information such as probability or certainty. Also, since the readings are getting updated at a fixed time interval, they should be timestamped to help determine the most recent reading. 
Another aspect of the system is security - only authorized personnel such as responders have access to the floor maps of the building. All these requirements call for a deeper understanding of modeling situations.

\begin{figure*}[t]
\centering
\includegraphics[width=150mm ,height=90mm]{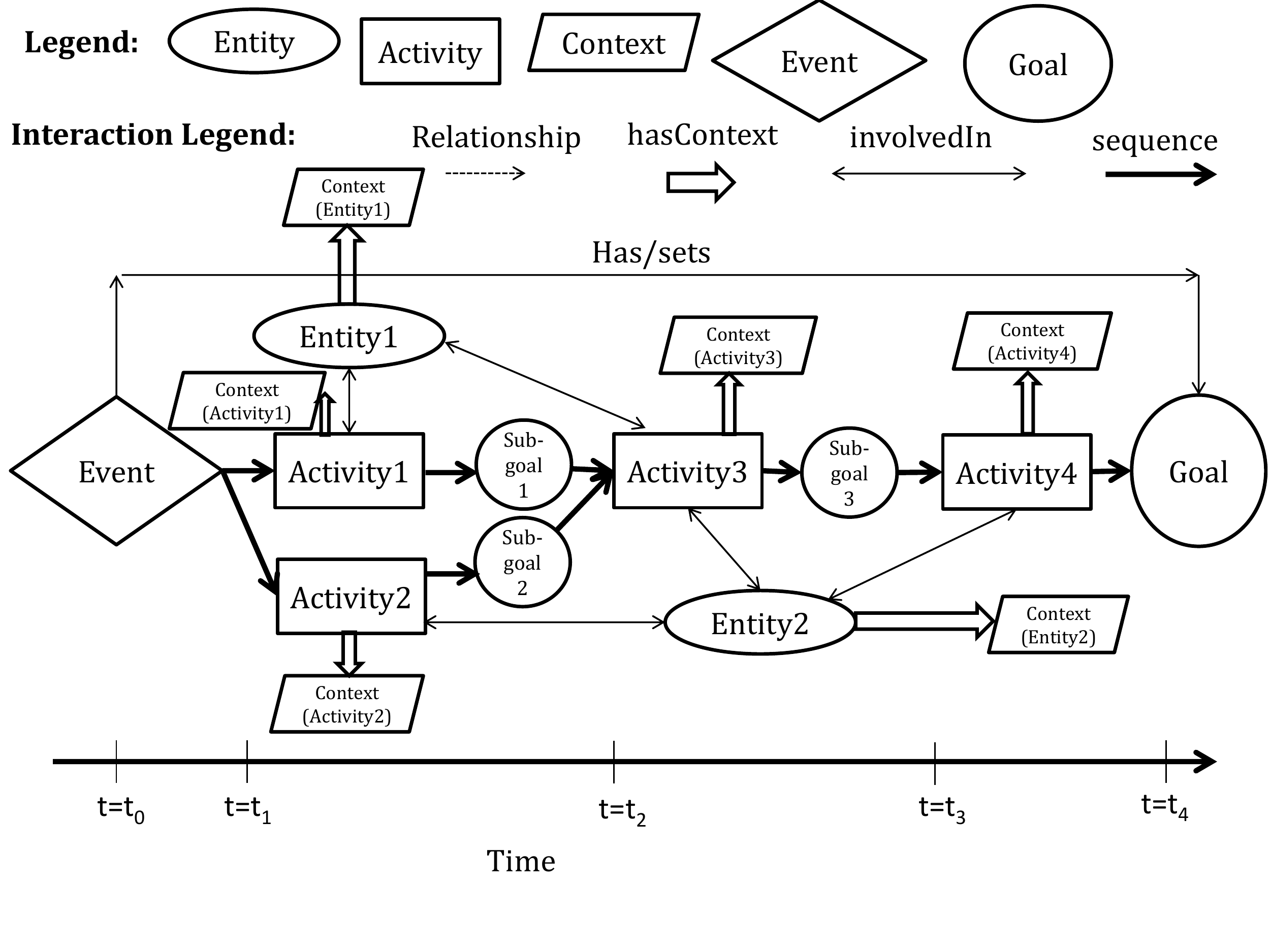}
\caption{Interaction between the primitives}
\label{primitives}
\end{figure*}

\section{Related Work in Context and Situation Modeling \label{sec:related}}

We briefly describe some of the existing context models and their underlying ontologies in this section, and examine their limitations.

CoBrA-Ont\cite{chen2003ontology} is a collection of ontologies for describing places, agents and events and their associated properties in an intelligent meeting-room domain. SOUPA\cite{chen2004soupa} was developed to provide pervasive computing developers with a shared and upper ontology that combines many useful vocabularies from different consensus ontologies such as FOAF, DAML-Time, RCC, BDI, and  Rei policy ontology. A full list of these well known ontologies can be found at \cite{semweb}.

Other contemporary ontologies include CONON\cite{gu2004ontology} where the context ontologies are divided into upper ontology and domain-specific ontologies; CoDAMoS\cite{preuveneers2004towards} where the context ontology is centered around four entities - user, environment, platform and service; ASC/CoOL\cite{strang2003cool} that enables context awareness and interoperability; Gaia \cite{ranganathan2003use} that incorporates ontologies for context awareness, service discovery and matchmaking, and interoperation between entities in a pervasive computing infrastructure mainly geared towards smart spaces; and GLOSS\cite{coutaz2010working} which employs ontologies for the precise understanding of various contexts and services in smart spaces.

Several surveys such as those by Reichle et al. (\cite{reichle2008comprehensive}, Krummenacher et al. \cite{krummenacher2007ontology} and Bettini et al. \cite{bettini2010survey}) have asserted that, of all the current ontologies used for context modeling, SOUPA is the most comprehensive ontology. However, both SOUPA and CoBrA-Ont have no provision for provenance, quality of context and multiple representations. CONON enables provenance by using the concept of sensed, derived, aggregated or deduced context but lacks features like comparability. Gaia takes on the challenge of modeling uncertainty and reasoning over it. However, their ontologies are restricted to the smart spaces domain. They do not model provenance either. A more detailed evaluation of all these ontologies can be found in the surveys mentioned. 

Moreover, a major shortcoming of general and exhaustive ontologies such as OpenCyc \cite{opencyc} is that they become too cumbersome to use in a system that is designed to be used efficiently and effectively in real time. Also, it is not possible for a small group of people to enumerate all the possible concepts and relationships between them that could be used in a practical mobile or desktop application or system. Hence, in our opinion, it is better to develop the base ontology and make it extensible for users just as most of the previous context models and ontologies have done. Our goal in this paper is to follow a similar approach and go one step further by addressing all the limitations that these existing ontologies posses.

We do not find much work in the literature where context-aware systems are extensively married to situational modeling. For modeling a situation (such as the Fire Incident mentioned earlier), none of the existing ontologies is adequate. This is mainly because these situations are rich and include events, which set a goal for the context-aware system to achieve, as well as entities interacting and performing a number of activities, along with their associated context. Moreover, a number of them have no provision for provenance, encoding bias and quality of context, and are often restricted to a single domain of use such as smart spaces. 

To address all these shortcomings, RoCoMO has been designed with an extensive ontological model-driven foundation along with capabilities to model both context and situation in a coherent and cohesive fashion. 

\begin{figure*}[t]
\centering
\includegraphics[width=150mm ,height=90mm]{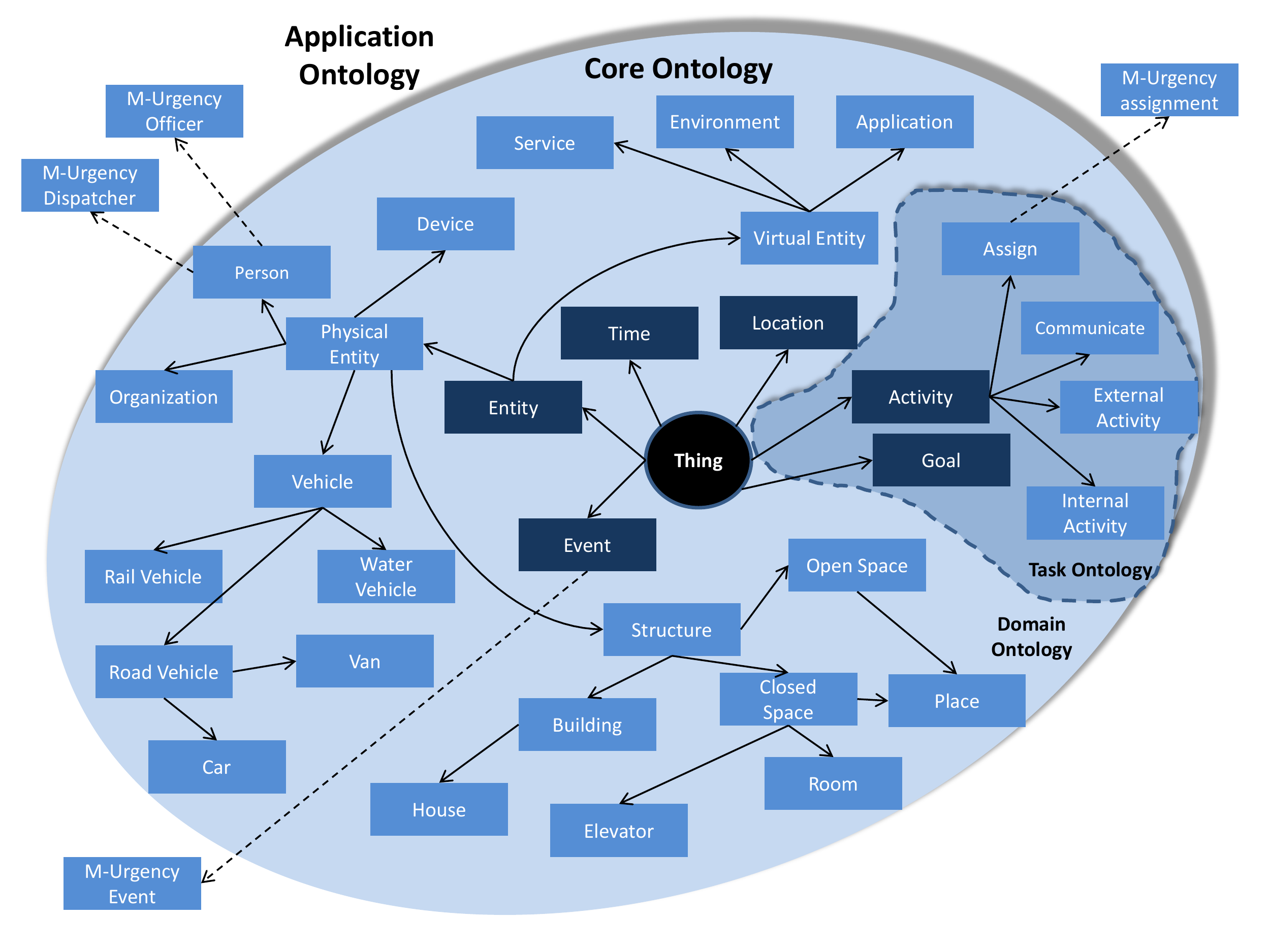}
\caption{Partial view of The RoCoM Ontology Version 1.0}
\label{rocomo}
\end{figure*}

\section{Structure, Design and Implementation of RoCoMO \label{sec:design}}

Figure \ref{primitives} shows how a situation can be represented in terms of the different concepts of RoCoMO - entities, events, activities and relationships. Each of them has contextual information associated with it such as location, identity etc.

At time t$_{0}$, the \emph{Event} catalyzes the context-aware system and sets the \emph{Goal} for it. To achieve the goal, different \emph{Entities} - Entity1 and Entity2 perform a sequence of \emph{Activities}, such as Activity1 and Activity2 beginning at time t$_{1}$. Every activity is driven by its own goal - which can be the overall goal that is set by the event, as in case of Activity4, or a sub-goal, as in case of Activity1. Every activity has a start time, an end time and a duration associated with it (as shown on the Time scale). It has a pre-condition and a post-condition which are goal(s) that should have been met before the activity starts and once the activity ends respectively. For instance, Activity3 starts at time t$_{2}$, when both SubGoal1 and SubGoal2 have been met, and ends at time t$_{3}$, when its own goal, SubGoal3, has been met. The duration for this activity is t$_{3}$ -  t$_{2}$. An activity can be atomic and non-interruptable.  

RoCoMO has been developed in OWL2 DL and has two components:
\begin{enumerate}
\item RoCoM Core Ontology which is divided into two upper level ontologies:
\begin{itemize}
\item RoCoM Domain ontology - This ontology includes concepts that characterize the knowledge of the domain i.e. the primitives \emph{Entities} and \emph{Events}, along with concepts that represent Location and Time. 
Examples of entities include persons, devices etc. while events can range from a simple service request to a road accident.
\item RoCoM Task ontology - This ontology characterizes the problem solving structure of the domain and provides primitives for describing the problem solving process i.e. \emph{Activities} as well as their \emph{Goals}. Examples of Activities include Calling, Scheduling etc.
\end{itemize}
\item RoCoM Application Ontology which has concepts that extend the core ontology concepts and are specific to an application.
\end{enumerate}

Figure \ref{rocomo} shows a partial view of the  RoCoM Core and Application Ontologies. 
The classes representing the primitives such as `entity', `event' and `activity', along with `location', `time' and `goal' are derived from the default OWL class `Thing' and form the top level classes in the RoCoM Core Ontology. Each of the classes derived from these top level classes represents a different, unique and unambiguous concept in the ontology. For instance, `physicalentity' is derived from `entity' and can be used to denote any entity that has a physical or logical form. It has more specific derived classes such as `person', `device' or `organization'. The fourth primitive `relationship' can be represented in OWL in many ways: between two classes (as a subclass/superclass), a class and an individual (as a member) or a specific relationship between two individuals (as object properties). 
The context of any element - entity, event or activity is represented using datatype properties in OWL. 

The following shows an OWL code snippet for the description of an individual instance named ``xyz" of class `person' which has a relationship with another entity (represented by the object property termed `daughter') and contextual information such as food preference (represented by the datatype property termed `likesfood').

\tiny
\begin{tabbing}
$\langle$Class \=rdf:about="\&person;person"$\rangle$\\
\>$\langle$ rdfs:label xml:lang="en"$\rangle$person$\langle$/rdfs:label$\rangle$\\
\>$\langle$rdfs:subClassOf rdf:resource="http://mind7.cs.umd.edu:8134/Rover/\\
\>physicalentity\#physicalentity"/$\rangle$\\
$\langle$/Class$\rangle$\\

$\langle$Object\=Property rdf:about="\&person;daughter"$\rangle$\\
\>$\langle$rdf:type rdf:resource="\&owl;FunctionalProperty"/$\rangle$\\
\>$\langle$rdf:type rdf:resource="\&owl;InverseFunctionalProperty"/$\rangle$\\
\>$\langle$rdfs:label xml:lang="en"$\rangle$daughter$\langle$/rdfs:label$\rangle$\\
\>$\langle$rdfs:subPropertyOf rdf:resource="\&person;contact"/$\rangle$\\
\>$\langle$inverseOf rdf:resource="\&person;father"/$\rangle$\\
\>$\langle$inverseOf rdf:resource="\&person;mother"/$\rangle$\\
$\langle$/ObjectProperty$\rangle$\\

$\langle$Data\=typeProperty rdf:about="\&person;likesfood"$\rangle$\\
\>$\langle$rdfs:label xml:lang="en"$\rangle$likesfood$\langle$/rdfs:label$\rangle$\\
\>$\langle$rdfs:subPropertyOf rdf:resource="\&person;likes"/$\rangle$\\
$\langle$/DatatypeProperty$\rangle$\\


$\langle$Nam\=edIndividual rdf:about=``\&person;xyz"$\rangle$\\
\>$\langle$rdf:type rdf:resource=``\&person;person"/$\rangle$\\
\>$\langle$rdfs:label xml:lang=``en"$\rangle$xyz$\langle$/rdfs:label$\rangle$\\
\>$\langle$person:likesfood rdf:datatype=``\&xsd;string"$\rangle$indian$\langle$/person:likesfood$\rangle$\\
$\langle$/NamedIndividual$\rangle$

\end{tabbing}
\normalsize

The core ontology can be further extended to concepts specific to an application, such as M-Urgency \cite{murgency}, to form a part of the Application Ontology. M-Urgency is a public safety application that enables mobile users to stream live video from their devices to local PSAP (Public Safety Answering Point) along with the audio stream, the real time location information and any personal and relevant information about the caller. 

Thus, a simple M-Urgency scenario can involve the entities (corresponding RoCoMO classes in parentheses): caller (`person'), dispatcher (`murgencydispatcher' extended from `person') and responder (`murgencyofficer' extended from `person'). For instance, because of an accident that is an event (`murgencyevent' extended from `event'), a series of activities follow such as, the caller calls the police, the dispatcher accepts the call, the dispatcher assigns (`murgencyassignment' extended from `Assign') a responder or officer to the call etc. This is only an illustration of how the concepts in the core ontology can be extended to model concepts specific to an application.


\section{Analysis and Evaluation of RoCoMO \label{sec:evaluation}}

Bettini et al. \cite{bettini2010survey} and Ye et al. \cite{ye2007} have specified a set of requirements that both context models and ontologies for pervasive computing environments should support. We assess RoCoM and RoCoMO on the basis of these criteria and explain how it addresses them:

\subsection{Representation of static and dynamic information} 

Contextual information can be static i.e. those aspects of a pervasive system that are invariant, such as a person's date of birth. However, the majority of contextual information is dynamic, such as location, with its persistence being highly variable. Every element of the RoCoM ontology, beginning with the top level classes like `entity', `activity' and `event', have their contextual information separated into two hierarchies - static and dynamic. This enables ease of distinction between contextual information that is persistent over a long period of time (static) and that which needs to be updated frequently based on its freshness (dynamic).

\subsection{Representation of temporal characteristics of primitives}

For every primitive such as an entity, activity or event, we have defined a time class that records properties such as its \emph{start time} - time at which the event/activity started or the entity came into being, \emph{end time} - time at which the event/activity ended or the entity ceased to exist (equivalent to the current time if the individual still exists), \emph{duration} or life time of an individual (difference of the start time and the end time) and \emph{recurrence} - frequency of repetition for an event/activity. The following OWL code snippet shows the time class:

\tiny
\begin{tabbing}
$\langle$Class \=rdf:about="http://mind7.cs.umd.edu:8134/Rover/time\#time"$\rangle$\\ \>$\langle$/Class$\rangle$\\

$\langle$owl \=:DatatypeProperty rdf:about="http://mind7.cs.umd.edu:8134/\\
\>Rover/time\#startTime"$\rangle$\\
\>$\langle$rdfs:domain rdf:resource="http://mind7.cs.umd.edu:8134/Rover/\\
\>time\#time/$\rangle$\\
\>$\langle$rdfs:range rdf:resource="\&xsd;dateTime"/$\rangle$\\
\>$\langle$rdfs:subPropertyOf rdf:resource="\&owl;topDataProperty"/$\rangle$\\
$\langle$/owl:DatatypeProperty$\rangle$\\

..........\\

$\langle$owl:DatatypeProperty rdf:about="http://mind7.cs.umd.edu:8134/Rover\\ \>/time\#repetition"$\rangle$\\
\> $\langle$rdfs:range:resource="\&xsd;string"/ $\rangle$\\
\> $\langle$rdfs:subPropertyOf rdf:resource="\&owl;topDataProperty"/ $\rangle$\\
$\langle$ /owl:DatatypeProperty $\rangle$

\end{tabbing}
\normalsize

\subsection{Timestamping} 

Timestamping the dynamic contextual information allows the system to determine the freshness and versioning of the contextual information which further enables resolution of conflicts and ambiguity. In RoCoMO, the contextual information is timestamped at two levels:
\begin{enumerate}
\item \emph{fine-grained level} - timestamping every dynamic contextual information of an individual instance of a primitive to keep track of when it was last modified and by which entity and 
\item \emph{coarse-grained level} - timestamping the individual instance itself to determine when it was modified and by which entity. 
\end{enumerate}

The following OWL code snippet shows how the \emph{hasMood} context of an individual \emph{xyz} of the \emph{person} class in RoCoMO is assigned a value \emph{happy} and is timestamped to determine when the contextual information was last updated.

\tiny
\begin{tabbing}
$\langle$Axi\=om$\rangle$\\
\>$\langle$annotatedTarget rdf:datatype="\&xsd;string"$\rangle$happy$\langle$/annotatedTarget$\rangle$\\
\>$\langle$rocomo-schema:timeStamp rdf:datatype="\&xsd;dateTime"$\rangle$\\
\>2012-09-18T14:00:00$\langle$/rocomo-schema:timeStamp$\rangle$\\
\>$\langle$annotatedProperty rdf:resource="\&person;hasMood"/$\rangle$\\
\>$\langle$annotatedSource rdf:resource="\&person;xyz"/$\rangle$\\
$\langle$/Axiom$\rangle$
\end{tabbing}
\normalsize

\subsection{Machine-interpretable representation of contextual information, Efficient context provisioning and Granularity of context}

The model and ontology must employ a machine-interpretable representation of context to tackle heterogeneity by using semantic annotations. These annotations can enable automatic exploitation and transformation of information in distributed context sharing scenarios as well as automatic context reasoning. They should provide efficient access paths to contextual information and represent it at different levels of abstraction. For instance, location of a user can be represented at a fine-grained level in terms of latitude/longitude and at a coarse-grained level in terms of the name of a city or a building.

RoCoMO is implemented in OWL2 DL which is expressive and allows more versatile knowledge representation. In OWL, context can be represented as annotated semantics via data properties and relationships between different elements can be represented via object properties. It also enables automatic context reasoning. Also, OWL represents information hierarchically which allows efficient provisioning of context and representation at multiple levels of abstraction or granularity.

\subsection{Encoding bias/ Comparability}

Contextual information sources constitute a variety of sensors and devices which often use different measurement and encoding systems, thus, resulting in a heterogeneous set of values describing the same entities. Hence, the context model and ontology must not depend on a particular symbol-level encoding, such as the representation of date in a particular format. It should provide means to compare and convert values with different scale and encodings. 

To address this, we annotate any measurable contextual information with an annotation property, called `scale', defined in a RoCoMO schema (This schema enables reification of every OWL statement that is part of RoCoMO). 
This removes the model's dependency on any particular encoding or measurement unit and also facilitates comparison or conversion from one unit to another. 

For instance, a person's context can include height which can be in feet, meters or any other unit. Thus, for person ``xyz", we can represent height and its measurement unit as:

\tiny
\begin{tabbing}
$\langle$Axi\=om$\rangle$\\
\>$\langle$annotatedTarget rdf:datatype="\&xsd;float"$\rangle$6.0$\langle$/annotatedTarget$\rangle$\\
\>$\langle$rocomo-schema:unit rdf:datatype="\&xsd;string"$\rangle$feet$\langle$/rocomo-schema:unit$\rangle$\\
\>$\langle$annotatedProperty rdf:resource="\&person2;height"/$\rangle$\\
\>$\langle$annotatedSource rdf:resource="\&person2;xyz"/$\rangle$\\
$\langle$/Axiom$\rangle$
\end{tabbing}
\normalsize

\subsection{Quality of Context (QoC)} 

Pervasive computing environments are highly dynamic and hence context data is characterized by properties such as incompleteness, ambiguity, uncertainty, inaccuracy, and temporal nature. For instance, in some environments, the contextual information may be incorrect due to a faulty sensor or incomplete due to lack of sufficient input. The model and ontology should be able to represent this imperfection.

Several papers including Gray and Salber \cite{gray2001modelling} introduced the notion of attaching information quality attributes to every piece of sensed context. To facilitate this, we have defined seven QoC attributes that model imperfection in contextual information - \emph{accuracy} to represent correctness, \emph{probability or confidence} to represent the certainty of being correct, \emph{coverage} to represent the range, \emph{resolution} to represent the smallest perceivable element, \emph{meanError} to represent average error,  and \emph{recurrence} to measure repeatability. These annotations are defined in the RoCoMO schema and can be attached to the appropriate contextual information or a relationship and can be propagated to applications. 

For instance, a person's context can include his/her weight which can be in kgs, pounds or any other scale. Also, the weight measure can have a mean error attached to it depending on the sensitivity of the instrument. 
Thus, person ``xyz" having weight 55 Kgs with an average error of 1 Kg, can be represented as:

\tiny
\begin{tabbing}
$\langle$Axi\=om$\rangle$\\
\>$\langle$annotatedTarget rdf:datatype="\&xsd;float"$\rangle$55.0$\langle$/annotatedTarget$\rangle$\\
\>$\langle$rocomo-schema:scale rdf:datatype="\&xsd;string"$\rangle$kgs$\langle$/rocomo-schema:scale$\rangle$\\
\>$\langle$rocomo-schema:meanError rdf:datatype="\&xsd;float"$\rangle$\\
\>1.0$\langle$/rocomo-schema:meanError$\rangle$\\
\>$\langle$annotatedProperty rdf:resource="\&person2;weight"/$\rangle$\\
\>$\langle$annotatedSource rdf:resource="\&person2;xyz"/$\rangle$\\
$\langle$/Axiom$\rangle$
\end{tabbing}
\normalsize

\subsection{Provenance and Traceability}

In order to provide adequate control and interpretation of contextual information, the model and ontology should provide the means to determine the source of data and transformations made to it. In RoCoMO, this is done at a coarse-grained level where we store, when the contextual information of any instance or an individual was created, when was it last modified, the last modification made to the instance and the entity by which it was made. However, we are not tracking every single modification made to every unique attribute or contextual information since this is too cumbersome at the modeling level. This can be achieved at the system level by logging context history and transformations. 

For instance, in our case study, we require the temperature of a room along with its source, its measurement scale, its time stamp and its certainty. Thus, the following OWL snippet represents an instance of an environment having temperature reading of 100 degree Fahrenheit, with certainty 0.9, created by a sensor instance `sensor1' at 2 pm on 09-18-2013.

\tiny
\begin{tabbing}
$\langle$owl\=:NamedIndividual rdf:about="\&environment;envreading1"$\rangle$\\
\>$\langle$rdf:type rdf:resource="\&environment;environment"/$\rangle$\\
\>$\langle$rdfs:label xml:lang="en"$\rangle$envreading1$\langle$/rdfs:label$\rangle$\\
\>$\langle$environment:temperature rdf:datatype="\&xsd;float"$\rangle$100.0$\langle$\\
\>/environment:temperature$\rangle$\\
\>$\langle$entity:createdBy rdf:resource="http://mind7.cs.umd.edu:8134/\\
\>Rover/sensor\#sensor1"/$\rangle$\\
$\langle$/owl:NamedIndividual$\rangle$\\

$\langle$Axi\=om$\rangle$\\
\>$\langle$rocomo-schema:probability rdf:datatype="\&xsd;float"$\rangle$0.9$\langle$\\
\>/rocomo-schema:probability$\rangle$\\
\>$\langle$rocomo-schema:timeStamp rdf:datatype="\&xsd;dateTime"$\rangle$\\
\>2013-09-18T14:00:00$\langle$/rocomo-schema:timeStamp$\rangle$\\
\>$\langle$owl:annotatedTarget rdf:datatype="\&xsd;float"$\rangle$100.0$\langle$/owl:annotatedTarget$\rangle$\\
\>$\langle$rocomo-schema:scale rdf:datatype="\&xsd;string"$\rangle$Fahrenheit\\
\>$\langle$/rocomo-schema:scale$\rangle$\\
\>$\langle$owl:annotatedSource rdf:resource="\&environment;envreading1"/$\rangle$\\
\>$\langle$owl:annotatedProperty rdf:resource="\&environment;temperature"/$\rangle$\\
$\langle$/owl:Axiom$\rangle$
\end{tabbing}
\normalsize

\subsection{Heterogeneity and Mobility} 

Pervasive computing environments are characterized by distribution, heterogeneity, unpredictability and unreliable communication links. 
Thus, the model and ontology should support these requirements and enable the aggregation and merging of the data when needed. RoCoM is an ontological model and promotes knowledge sharing and reuse across distributed systems and applications in pervasive computing environments. Hence, even if the sources of context are heterogeneous, distributed and partitioned, the contextual information can be shared and aggregated across environments.

\subsection{Ease of development}

RoCoMO is developed on the principle of Model-Driven Development. The ontology is also available publicly. Hence, developers have adequate support for development and implementation.

\subsection{Flexibility, extensibility, applicability, generality, evolvability and completeness}

Context models and ontologies should not be rigid but flexible and extensible. Thus, they should not be restricted to a single domain, and should be able to support new and varied application domains. 
They should evolve with the applications and their context needs. 

RoCoMO is structured in a modular fashion with clear distinction between the Core and Application ontologies. Also, as the applications evolve, more concepts can be added to it. Thus, it is easily extensible, flexible and evolvable. It does not target any specific domain in pervasive computing and is intended to be general and applicable across several applications and domains. As a result, we do not claim that the ontology is complete.

\subsection{Interoperability} 

Since several existing projects use standard upper ontologies, every generic ontology should be interoperable i.e. its term definitions must be consistent with other standard, generic and consensus ontologies such as SOUPA \cite{chen2004soupa}. 
This also enables reuse of domain knowledge \cite{gu2004ontology}. 

We have designed RoCoMO to be interoperable with other ontologies, for instance SOUPA\cite{chen2004soupa}, via the \emph{equivalentClass} and \emph{equivalentProperty} OWL statements. 
The example below shows that the RoCoMO \emph{person} class is defined equivalent to the \emph{person} class in SOUPA and the \emph{dateofbirth} property is defined equivalent to \emph{birthDate} property in SOUPA.

\tiny
\begin{tabbing}
$\langle$Class \= rdf:about="\&person;person"$\rangle$ \\
\>$\langle$rdfs:label xml:lang="en"$\rangle$person$\langle$/rdfs:label$\rangle$\\
\>$\langle$equivalentClass rdf:resource="http://pervasive.semanticweb.org/ont/2004\\
\>/06/person\#person"/$\rangle$\\
\>$\langle$rdfs:subClassOf rdf:resource="http://mind7.cs.umd.edu:8134/Rover/\\
\>physicalentity\#physicalEntity"/$\rangle$\\
$\langle$/Class$\rangle$ \\

$\langle$DataP\=roperty rdf:about="\&person;dateofbirth"$\rangle$ \\
\>$\langle$rdfs:label xml:lang="en"$\rangle$dateofbirth$\langle$rdfs:label$\rangle$ \\
\>$\langle$rdfs:subClassOf rdf:resource="\&person;personalinfo"/$\rangle$\\
\>$\langle$equivalentProperty rdf:resource="http://pervasive.semanticweb.org/ont/\\
\>2004/06/person\#birthDate"/$\rangle$\\
$\langle$/DataProperty$\rangle$

\end{tabbing}
\normalsize

\begin{figure*}[t]
\centering
\includegraphics[width=150mm ,height=90mm]{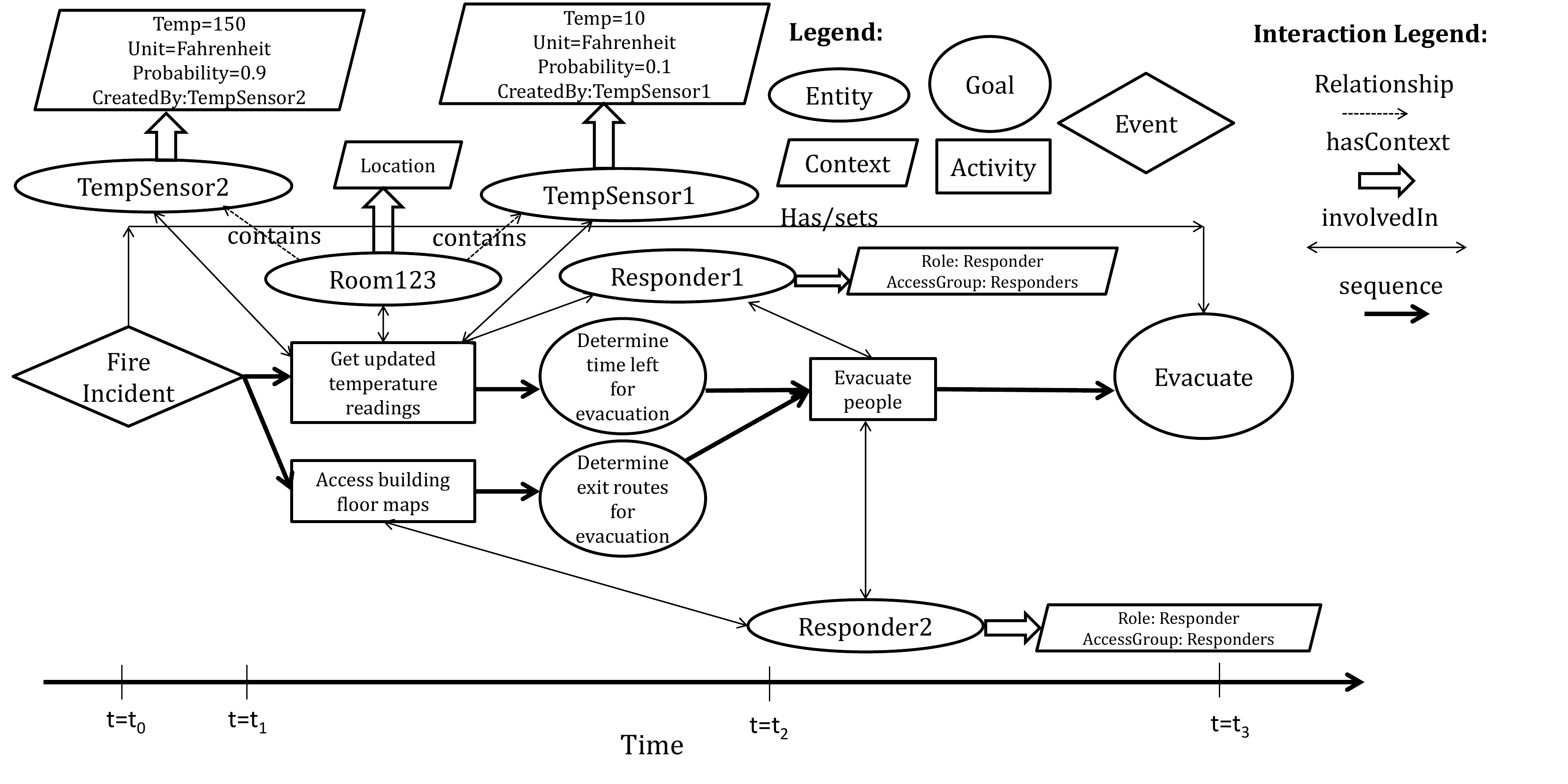}
\caption{Fire Incident Case Study modeled using RoCoMO}
\label{casestudy}
\end{figure*}

\subsection{Clarity, Coherence, Redundancy and Orthogonality} 
The concepts in RoCoMO are unique, unambiguous, independent and consistent.

\subsection{Security and Privacy}

These are implemented in RoCoM using Role Based Access Control (RBAC) for groups and members. A group is an instance of type `accessgroup' class. This class has object properties like `groupmember' which includes the entities like users or devices that can be assigned to an instance of the group. The `accessgroup' class also has an object property called `privileges' which defines the permissions that the group can have. These permissions can be in the form of an `activity' that the group is allowed to perform. Every entity can belong to multiple access groups while each access group can have multiple entities and privileges. This form of access control, obtained by assigning users to groups and granting privileges to groups rather than individual users, reduces the number of associations involved that need to be managed. Hence, it is easier to define security policies around this framework.

\section {Modeling the Fire Incident Case Study using RoCoMO \label{sec:usecases}}
In this section, we revisit the Fire Incident case study from Section \ref{sec:scenario} and illustrate how RoCoMO can be used to model it. Figure \ref{casestudy} shows a graphical representation of the situation modeled in RoCoMO. At time t$_{0}$, the \emph{Fire Incident} Event triggers the situation that follows and sets the Goal \emph{Evacuate}. This goal can be subdivided into smaller sub-goals which can be performed by one or more activities. Entity \emph{Responder1} performs the Activity \emph{Get updated temperature readings}, at time t$_{1}$, to get the context information of room \emph{Room123} - the updated temperature readings from the temperature sensors \emph{TempSensor1} and \emph{TempSensor2}. The Goal for this activity is \emph{Determine time left for evacuation}. Since the information is timestamped, the system can refresh it periodically based on its freshness and this resolves any ambiguity. 

The contextual information also has a probability measure attached to it. As shown in the figure, the probability of temperature reading, from \emph{TempSensor1}, being correct is 0.1 which means it is highly unreliable and that the sensor could be faulty. This is evident by the fact that it shows a reading of 10 deg Fahrenheit while the other sensor shows a reading of 150 deg Fahrenheit. Also, the contextual information has source or provenance information attached to it and so this determines which reading came from which sensor. Based on the temperature reading and its encoding (Fahrenheit in this case), the system can calculate how much time it will take till the temperature reaches the value at which the building bursts into flames. This is the amount of time that the responders have for evacuation. 

Simultaneously, another entity \emph{Responder2} is performing the activity \emph{Access building floormaps} with the Goal - \emph{Determine exit routes for evacuation}. Since the responders belong to the AccessGroup \emph{responders}, the system checks their privileges (which are inherited from the `responders' accessgroup) and grants access to the temperature readings from the sensor and building floor plans. Once these two activities have achieved their goals, both the responders start the activity \emph{Evacuate people}, at time t$_{2}$, and achieve the goal set by the event.

\section {Conclusion and Future Work \label{sec:conclusion}}

In this paper, we described a generic, flexible and extensible ontology called Rover Context Model Ontology(RoCoMO) and illustrated its benefits for context and situation modeling in pervasive computing environments, via a practical case study. We highlighted several shortcomings of contemporary context models and ontologies and explained how RoCoMO addresses them. We also established the utility of its capabilities by evaluating it against several criteria that ontologies for context and situation modeling should possess. Our next step is to develop a GUI to allow users to browse and explore the RoCoM ontologies, and further extend and modify them. An API for working with the ontology will also be available. Our aim is to encourage users to build applications and systems using this ontology so that they can be used in both in an isolated manner or integrated with Rover II.

\bibliographystyle{IEEEtran}      
\bibliography{rocomo}

\end{document}